\documentclass[aps,prl,preprint,superscriptaddress]{revtex4}
\usepackage{amssymb,amsmath,amsfonts,graphicx,color}
\usepackage{epsfig,color,times}
\usepackage{amsmath, mathrsfs}
\begin{document}
\title{Soft granular matter}
\author{Serge Mora}
\email[]{smora@univ-montp2.fr}
\affiliation{Laboratoire de M\'ecanique et de G\'enie Civil de Montpellier. UMR 5508, Universit\'e Montpellier 2 and CNRS. Place Eug\`ene Bataillon. F-34095 Montpellier Cedex, France.}
\author{Yves Pomeau}
\email[]{pomeau@lps.ens.fr}
\affiliation{University of Arizona, Department of Mathematics, Tucson, AZ, USA.}
\date{\today}
\begin{abstract}
We consider a dilute system of small hard beads or hard fibers immersed in a very soft gel able to withstand large elastic deformations. Because of its low to very low shear modulus, this system is very sensitive to small forces.  We calculate the local deformation induced by a constant volume force, the inclusion weight. We explain how this deformation could be put in evidence by using techniques similar to the PIV method (particle image velocimetry) used to show complex velocity fields in transparent fluids.  
\end{abstract}

\maketitle

\section{Introduction}

This paper is to introduce a new kind of material. We call it "soft granular matter" because it is made of grains (beads or fibers of hard solid, basically non deformable in the conditions we shall consider) immersed in a very soft elastic and transparent solid which is able to withstand large deformations. Hard beads immersed in a gel have been considered before by Chaudhury and collaborators \cite{Chaud} in the case of beads entering the gel from its free surface under the influence of gravity. In this case, surface tension plays also an important role, which is a priori not the case in the situation we have in mind, namely dilute beads immersed in the gel before the gelification.  

We outline first some expected properties of this material and then how it can be used to investigate various physical problems. 

The mechanical properties of very soft gels have been subjected lately \cite{nous,d'autres} to detailed studies because, contrary to often believed, they make excellent elastic solids, being able  to stand reversibly large strains (contrary to most materials we think as elastic, like iron or glass). Those gels have been shown recently \cite{nous} to be sensitive to small forces like surface tension. We have shown the existence of a Rayleigh-Plateau instability for the first time in this kind of (soft) solid. Therefore it is an important issue to look at the possibility to use this soft matter to measure small forces by their deformation. This is a priori a non trivial task, because gels are made almost uniquely of water, so that the elastic stress in the gel is almost invisible in its atomic structure. A  way to overcome this difficulty is to immersed beads into a gel so that they follow the distortions of the gel. We explain at the end possible ways to measure such distortions. 

\section{The dilute regime}

\begin{figure}[!h]
\includegraphics[width=0.2\textwidth]{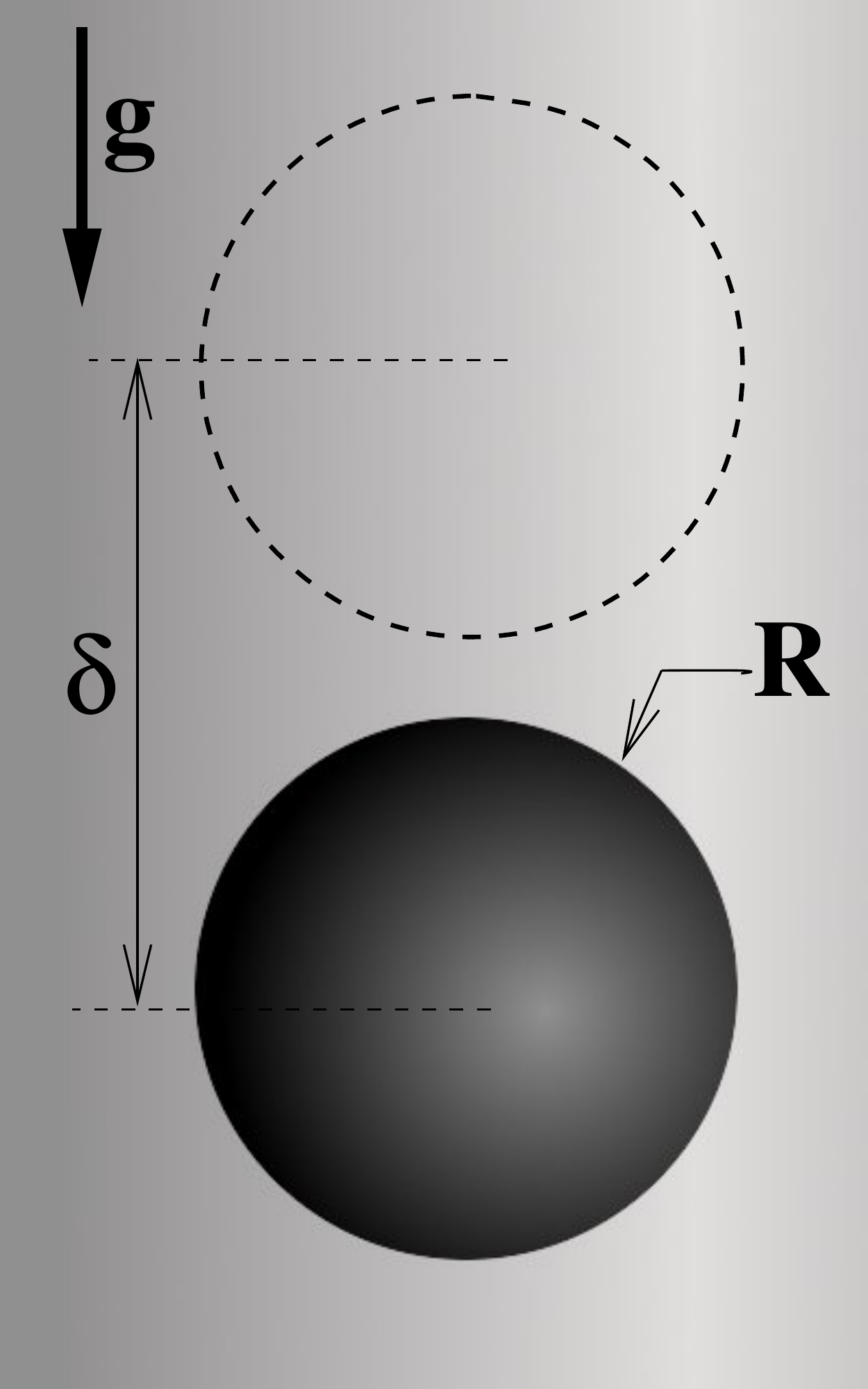}
\caption{A bead of radius $R$ is placed in a gel of shear modulus $\mu$ at zero gravity. Once the gravity is turned on, the vertical displacement of the bead under its own weight is $\delta$. }
\label{fig : gravity}
\end{figure}

\begin{figure}[!h]
\includegraphics[width=0.5\textwidth]{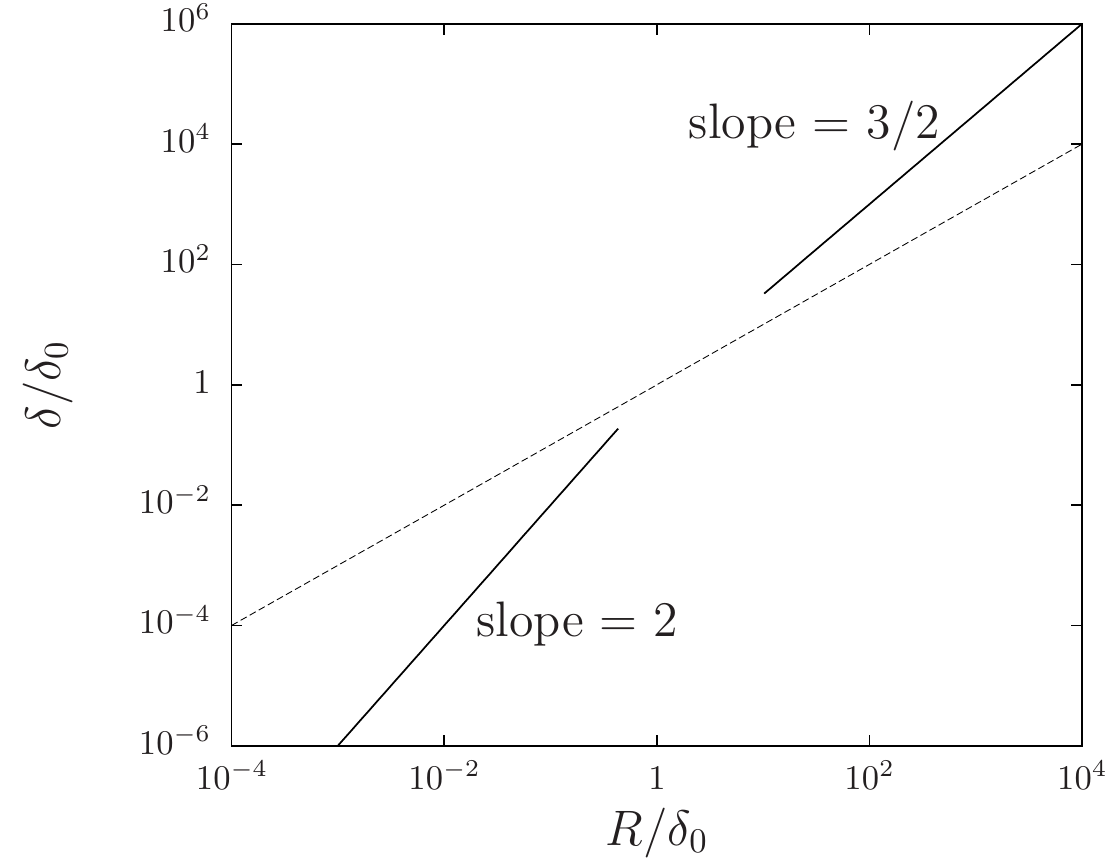}
\caption{Asymptotic behaviours of the dimensionless downshift as a function of the dimensionless radius. The dotted line, representing slope 1, is a guide for the eyes.}
\label{fig : sphere}
\end{figure}

Consider a suspension of (hard) beads in a soft gel. Because the gel is soft, we have to consider first  the tendency of beads in real life to fall down under their own weight. Let $\rho$ be their relative density with respect to the gel (this is the difference between the mass density of the beads and of the gel, assumed to be positive), let  $R$ be the radius of the beads (assumed to be spherical although this is not a crucial point if the small bead is compact with all typical length of the same order). Thin needles instead of compact beads are also interesting and are considered below, after compact beads. Let $\mu$ be the shear modulus of the gel. Let $\delta$ be the vertical displacement of the immersed bead under its own weight (Fig. \ref{fig : gravity}). One must compare $\delta$ and the radius  $R$ of the bead. Assume first that $\delta \gg R$, and let us find its order of magnitude. This is derived, as usual in mechanical problems of this kind, by comparing two energies: the potential energy of the bead in the field of gravitation and the elastic energy due to the downshift of the bead in the elastic medium. We assume that the gel is neo-Hookean, a fair representation of many soft gels. In  neo-Hookean material, the elastic energy of a strained configuration is 
$\mu$ (shear modulus) times the deformed volume times the square of the strain. The strain is of order 1, because there is no length scale in the equations of neo-Hookean materials, so that the gradient of a displacement field (the strain) has to be of order one. Therefore the elastic energy is of order of magnitude $\mu \delta^3$ because the domain where there is deformation is of size $\delta$, the only length scale, as the radius of the bead can be neglected. At equilibrium this elastic energy must be of the same order of magnitude as the drop of potential energy in the gravity field, namely $\rho R^3 g \delta$. This gives 
$$\delta \sim  R^{3/2} \delta_0^{-1/2},
$$ where $\delta_0 = \frac{\mu}{\rho g}$. This length gives the magnitude of the vertical fall if the radius $R$ is much bigger than $\delta_0$. \\
Let us consider now the opposite limit $R \ll  \delta_0$, a limit where we expect that $\delta \ll R$ and so the deformation to be small enough to make the Hookean (instead of neo-Hookean before) approximation of a small strain valid. Let us estimate $\delta$ in this case. The change in the gravitation energy is still 
$\rho g R^3 \delta$. The Hookean elastic energy of deformation is  $\mu$ times the square of the strain times $R^3$ (volume of the bead, also the order of magnitude of the volume where the elastic energy is stored: because the equations of linear Hookean elasticity are elliptic the range of the perturbed domain is fixed by the boundary conditions, implying therefore the radius of the bead). The order of magnitude of the dimensionless strain is $\delta/R$ so that  the balance of the elastic and gravitation energy yields 
$$\delta \sim  \frac{R^2}{\delta_0}$$ 
in this case, a result valid in the limit $R \ll \delta_0$.  Of course, this last result is consistant with the exact calculation. The Navier-Cauchy equation for a linear elastic and incompressible medium is $\nabla \cdot \underline{\sigma}+\nabla p=0$ with the incompressibility condition $\nabla {\bf u}=0$, where ${\bf u}$ is the displacement field, $\underline{\sigma}$ the extra-stress tensor ($\sigma_{ij}=\mu(u_{i,j}+u_{j,i})$ in Cartesian coordinates) and $p$ the (non-hydrostatic) pressure. The boundary condition at the bead surface is $u=\delta$. These equations are equivalent to those for the drag force exerted on a spherical bead with very small Reynolds numbers in a viscous fluid, changing the displacement field with the velocity field, the downshift $\delta$ with the bead velocity, and $\mu$ with the fluid viscosity. Stokes'law  \cite{Stokes} gives  $F=6\pi R \mu \delta$ with $F=\frac{4}{3}\pi R^3 \rho g$ and then $\delta=\frac{2}{9}\frac{\rho g}{\mu}R^2=\frac{2}{9}\delta_0R^2$.\\        
The asymptotic behaviours of $\delta$ are shown in Fig. \ref{fig : sphere}.\\

Let us consider now a long needle of radius $R$ and length $L$, such that $L \gg R$, with a relative mass density $\rho$. The other quantities have the same definition as before. In all cases, let $\delta$ be the downshift due to gravity when the needle is left in gel of shear modulus $\mu$. Consider first a very long needle. The drop of gravitation energy due to the downshift is of order $\rho R^2 L g \delta$. Assuming $\delta \gg R$ and that the strain is of order one, one finds that the elastic energy due to the deformation of the gel is of order $\mu L \delta^2$. Balancing the two enegies, one obtains 
$$ \delta \sim \frac{R^2}{\delta_0} \mathrm{.}$$ 
This is derived under the condition $\delta \gg R$. In the opposite limit we use the previous correspondence relation between the linear elastic problem and the Newtonian fluid problem. 
The drag force exerted on a fiber with the relative velocity $V$ in a fluid (viscosity $\eta$) is $F= \frac{2\alpha\pi}{\log(L/R)}\eta V L$ where $\alpha=1$ for a velocity parallel to the fiber axis, and $\alpha=2$ for a velocity perpendicular to it.  This is valid only in the limit of low Reynolds numbers \cite{Lamb}. Using the same changes as for beads, one obtains $\delta = \frac{R^2}{\alpha \delta_0}\log\frac{L}{R}$. The order of magnitude of $\delta$ is the same as in the opposite case $\delta \gg R$, which is consistent with the condition $ R\ll \delta_0$. \\ 


Let us return to the case $\delta \gg R$, but in the limit $ R \ll L \ll \delta$, although we had before $L\gg \delta$. The gravitational energy due to the downshift is still $\rho R^2 L g \delta$. The elastic energy of deformation is just $\mu \delta^3$ because the pertubed region in the gel is the largest length, namely $\delta$. The balance of the two energies yields $\delta \sim \left(\frac{L R^2}{\delta_0}\right)^{1/2}$, valid if $R \ll L\ll \frac{R^2}{\delta_0}$.  

\section{The semi-dilute regime}

For a uniformly distributed random suspension Batchelor \cite{Batchelor} showed that the average sedimentation speed of monodisperse spherical particles (radius $R$) through a viscous fluid (viscosity $\eta$) under the effect of gravity is, at small Reynolds Number $v= \frac{2}{9}\frac{\rho g}{\eta}R^2(1-6.55\phi+{\cal O}(\phi^2))$ where $\phi$ is the suspension volume fraction. Considering the elastic problem, we deduce that, under the condition  $\delta \ll R$, the average displacement of the beads is :
\begin{equation}
\delta=\frac{2}{9}\frac{\rho g}{\mu}R^2\left(1-6.55\phi+{\cal O}(\phi^2)\right),
\end{equation} 
$\phi$ being the beads volume fraction.

\section{Applications}

To have a dispersion homogeneous at the scale of the gel,  the size system must be much bigger than the length scale $\delta$, otherwise, the immersed particles will tend to fall to the bottom of the gel. As those gels are made from a liquid solution, one should prepare it by putting also the inclusions initially in the liquid destined to make the gel. To avoid the settling of the particules to the bottom of the solution during the gelification, on may think to rotate this solution around an horizontal axis. 

What is the interest of this material? As said above, the gels made in this way are transparents. Therefore the immersed particles will be like the small particles used in the visualisation of fluid flows, in the now well used technique of PIV (particles image velocimetry) \cite{PIV}. There are some variants of this technique, but the general idea is always the same: by taking images of the particles at slightly different times and by assuming that they are convected by the local fluid velocity, the field of the displacements between two snapshots gives the velocity field, after division by the interval of time. In the case of the particles immersed in a gel, the two snapshots would be taken at different values of an imposed external stress and so give the distribution  of strain in the gel. No such information is usually available in regular elastic solids where the strain can at best be inferred from displacements on the surface. Photoelasticimetry  is a possible way of measuring bulk deformation in elastic solids, but the sensitivity of this method has a quite weak.   

This device could be used in fondamental problems of solid mechanics, like for instance the distribution of strain near the tip of fractures, already studied, although indirectly, in gels \cite{montpellier}. Without having done experiments and particularly  without having an idea of the ultimate sensitivity of the method, it is difficult to discuss other applications. We can only mention several problems where small forces need to be measured. Such a fundamental problem  is the so-called Abraham-Minkowski dilemma concerning radiation pressure in refractive media \cite{abraham}, still a controversial matter after more than a century. Of course one may think also to measurements of the weak force of gravitation: could the particles in a gel be moved measurably by the Newtonian attraction of an external mass? If it can, it could be a way to measure how Netwon's law changes, if it does, at small distances.   

Another field of use of this material could be the study of its behaviour under large strains: gels are well approximated by the neo-Hookean theory. So one might wonder if they change behaviour if the density of beads becomes finite (instead of small as in the applications we thought about). It could be that the randomness so introduced yields regions of large strain, even with a finite average strain. Such a concentration of strain  and stress is believed to be responsible of the irreversible creeping in some materials.  It could be that, locally, because of the fluctuations a Biot-like instability \cite{biot} occurs at small scale, inducing so an irreversible transformation of the material.

{}
 \ifx\mainismaster\UnDef%
 \end{document}
  \fi